# Increased Phase Coherence Length in a Porous Topological Insulator

Alex Nguyen[1,2]†, Golrokh Akhgar[1,2]*†, David L. Cortie[2,3,4], Abdulhakim Bake[3], Zeljko Pastuovic[4], Weiyao Zhao[1,2], Chang Liu[1,2], Yi-Hsun Chen[2,5], Kiyonori Suzuki[1], Michael S. Fuhrer[2,5], Dimitrie Culcer[2,6], Alexander R. Hamilton[2,6], Mark T. Edmonds[2,5] Julie Karel[1,2]*

[1]Department of Materials Science and Engineering, Monash University, Clayton, 3800, Victoria, Australia.
[2]ARC Centre of Excellence in Future Low-Energy Electronics Technology, Monash University, Clayton, 3800, Victoria, Australia.
[3]Institute for Superconducting and Electronic Materials, University of Wollongong, North Wollongong, 2522, New South Wales, Australia.
[4]Australian Neutron Science and Technology Organisation, New Illawarra Road, Lucas Heights NSW 2234, New South Wales, Australia.
[5]School of Physics and Astronomy, Monash University, Clayton, 3800, Victoria, Australia.
[6]School of Physics, The University of New South Wales, Sydney, 2052, New south Wales, Australia.

†These authors contributed to this work.
*Corresponding authors: gol.akhgar@monash.edu, julie.karel@monash.edu

**Abstract**

The surface area of $Bi_2Te_3$ thin films was increased by introducing nanoscale porosity. Temperature dependent resistivity and magnetotransport measurements were conducted both on as-grown and porous samples (23 and 70 nm). The longitudinal resistivity of the porous samples became more metallic, indicating the increased surface area resulted in transport that was more surface-like. Weak antilocalization (WAL) was present in all samples, and remarkably the phase coherence length doubled in the porous samples. This increase is likely due to the large Fermi velocity of the Dirac surface states. Our results show that the introduction of nanoporosity does not destroy the topological surface states but rather enhances them, making these nanostructured materials promising for low energy electronics, spintronics and thermoelectrics.

**Introduction**

Bi$_2$Te$_3$ is a three-dimensional topological insulator (TI) characterized by an insulating bulk and topological metallic surface states.[1-4] The electronic structure of the surface states exhibits a linear dispersion relation and spin-momentum locking. Additionally, the surface states are topologically protected from backscattering, and are therefore appealing for low energy electronics since dissipationless currents can be achieved.[3, 5-7] Owing to the unique electronic properties and large spin-orbit coupling, 3D TIs also exhibit efficient charge to spin current conversion arising from the spin-momentum locking in the surface states.[8, 9] The largest reported spin orbit torque[10] has been found using a 3D TI, making this class of materials particularly promising for future spintronic devices.[11-13]

Magnetotransport experiments have been extensively utilized to probe the topological surface states. In TIs, these states exhibit weak antilocalization (WAL) in their magnetoconductivity due to spin-momentum locking. According to the theory, the charge carriers move in $2\pi$ circles above the Dirac point and the $\pi$-Berry phase associated with these helical surface states changes the interference of the incoming and outgoing time-reversal paths from constructive to destructive.[14-16] The quantum correction to the conductivity of the 2D systems with strong spin-orbit coupling was initially derived by Hikami, Larkin and Nagaoka (HLN) [17] and Bergmann described the intuitive picture of WAL[18, 19]. In the case of surface states in TIs, the WAL correction is similar to the correction described by HLN as the two systems are expected to be in the same universality class of 2D systems. However, in contrast to other 2D systems, there is no crossover from WAL to weak localisation (WL) in TIs with increasing magnetic fields as the spins in the Dirac surface states of the TI are perfectly locked to momentum.[16] Typically 2D magnetoconductivity can be described using the HLN formulation in order to extract information about the nature of the transport (e.g. phase coherence length, decoherence mechanisms, number of conducting channels, 2D versus 3D scattering, etc.).[17, 18] Studies have shown the characteristics of the quantum transport can be controlled through, for instance, gating, doping, film thickness, a capping layer, or surface impurities.[20-26] Despite the fact that the phenomena being probed are present on the surface, the vast majority of these studies were conducted with a constant surface to volume ratio.

A conventional path to increase the surface to volume ratio is through the introduction of porosity.[27] This will typically also increase the disorder in the material; therefore, most work has avoided porous materials. Research has traditionally been directed towards studying highly ordered crystals with very low defect densities and impurity concentrations.[21, 28-30] However, recent theoretical work has predicted the existence of topological conductive surface states and insulating bulk states in amorphous materials.[31-33] These theoretical models concluded that in the bulk of the material there are energies that are not eigenvalues of the Hamiltonian, i.e. there is a bandgap in the bulk. By contrast, on the surface where the amorphous solid interfaces with a vacuum, these energy values become eigenvalues of the Hamiltonian. The study conducted by Costa et al. in particular is notable for modelling the density of states in amorphous bismuthene instead of a generic amorphous material.[33] Moreover, an exciting recent study has found that a topological phase transition from trivial insulator to TI can be induced by making a topologically trivial crystalline material either disordered[34] or amorphous.[35] These theoretical studies point to exciting possibilities to discover topological properties in disordered materials, a prospect which has been largely unexplored experimentally.

In this work, we introduce nanoscale porosity in Bi$_2$Te$_3$ in order to modify the surface to volume ratio. We will show that increasing the surface area, while also increasing disorder, leads to an enhancement in the WAL effect and, remarkably, a doubling of the phase coherence length. We suggest that the larger phase coherence length in the porous samples arises from an increased fraction of surface states, which have a higher velocity, contributing to the magnetoresistance.

**Results**

Nanocrystalline $Bi_2Te_3$ thin films were grown through a shadow mask using molecular beam epitaxy (Figure 1a). High resolution transmission electron microscopy (HRTM) measurements on the as-grown films indicated a nanocrystalline structure, with an average crystal size of ~20nm. Figure 1b displays high angle annular dark field (HAADF) cross-sectional TEM of the ion-irradiated 70 nm thin film. Nanoscale porosity can be observed throughout the entire film; the majority of the pores (99%) are between 1 and 10 nm in size. Upon examination of a 2D cross sectional TEM image, approximately 22% of the area contained voids (see Supplementary Information). Figure 1c depicts a bright field image of the same sample at higher magnification. The Energy-dispersive X-ray spectroscopy (EDS) measurement confirms that Ne ions were able to penetrate the entire thickness of the thin film which created porosity and left cavities throughout the whole material. EDS measurements show that no Ne ions were detected after ion irradiation and the sample is slightly Te poor (see Supplementary Information).

Figure 2a-b show the normalised magnetoresistance (($\rho_{xx}(B) - \rho_{xx}(0))/\rho_{xx}(B=0)$) as a function of applied magnetic field for the 23 nm as-grown and porous samples at various temperatures (see Supplementary Information for 70 nm samples). These curves are representative of all samples measured and show positive magnetoresistance due to WAL. In TIs, this quantum coherent transport arises due to spin momentum locking of the topological surface states.[14-16] The WAL correction occurs at sufficiently low temperatures, and in these samples is not present above ~12K, at which point the magnetoresistance is parabolic. Figure 2c-d compares the magnetoresistance at 2K for the porous and as-grown samples of different thicknesses. The WAL effect is more pronounced at 2 K in the porous sample compared to the as-grown sample for both sample thicknesses. Remarkably, Figure 2c-d shows that the introduction of porosity has a significant effect on the magnetotransport and actually strengthens the WAL effect. The low field cusp is much stronger in the porous samples for both thicknesses, and this observation is true for all temperatures up to 10 K (see Supplementary Information). The WAL interference effect is reduced with increasing magnetic field as a result of the additional phase introduced by the applied field. Therefore, in order to analyze the 2D WAL data accurately, the quantum transport correction should be limited to low fields (B < 0.5 T), and the HLN 2D localisation theory is derived for the transport correction in low magnetic field only.[17, 18]

To probe the origins of the quantum transport, the 2D magnetoconductivity data has been fit using the reduced HLN formula:[17, 18]

$$\Delta \sigma_{xx} = -\frac{\alpha e^2}{\pi h}\left[\ln\left(\frac{\hbar}{4eBL_\phi^2}\right) - \Psi\left(\frac{1}{2} + \frac{\hbar}{4eBL_\phi^2}\right)\right] \quad (1)$$

Where $B$ is the applied magnetic field perpendicular to the sample, $\Psi$ is the digamma function, e is an elementary charge, $L_\phi$ is the phase coherence length and $\alpha$ is a coefficient relating to the conduction channels.[17] Figure 3a,b shows a representative fit for the porous 23 nm and 70 nm film up to 0.5 T (see Supplementary Information for other samples). The pre-factor $\alpha$ provides information about the channels contributing to the conduction; each conducting channel contributes -0.5. WAL is a competition between phase decoherence and surface to bulk scattering. If the phase coherence time is much larger than the surface to bulk scattering time, substantial surface-bulk scattering will occur and the sample will have a single transport channel with $\alpha = -0.5$. On the other hand, if the surface to bulk scattering time is large in comparison to the phase decoherence time, the top and bottom surfaces will be decoupled and each contribute -0.5 to the conduction, resulting in $\alpha = -1.0$.[20-22, 25, 26, 36, 37] In the 23 nm samples, $\alpha = -0.5$, indicating the surface and bulk are coupled and behave as a single transport channel. By contrast, the 70nm samples show $\alpha = -1.0$, meaning the top and bottom surfaces are decoupled; the increase in thickness likely results in a more significant effect of the bulk conductivity which reduces the coupling between the surfaces.

Figure 3c shows the temperature dependence of the phase coherence length ($L_\phi$) extracted from the reduced HLN fits. Depending on the dimensionality of the system, $L_\phi$ follows a power law scaling $L_\phi = T^{-\beta}$ with $\beta$=0.5 for 2D Nyquist dephasing and 0.75 for 3D Nyquist dephasing.[38-42] Surprisingly, for both sample thicknesses the phase coherence length approximately doubles with the introduction of porosity.

Figure 4a shows the longitudinal resistivity ($\rho_{xx}$) as a function of temperature for the 23 and 70 nm as-grown and porous films. After ion irradiation, $\rho_{xx}$ for a given thickness surprisingly decreases, indicating the samples become more metallic. The charge carrier densities for these samples, as determined by Hall effect measurements, ranged from 4.8 x $10^{20}$ to 1.5 x $10^{21}$ cm$^{-3}$; all samples except the 70nm porous sample were p-type. (See Supplementary Information) Given the porous samples are Te poor, n-type behaviour would be expected for the 23nm porous sample as well. The relatively high carrier densities and the unexpected sign of the carriers is likely due to charge puddling; a further analysis of the MR behaviour confirmed this result. (see Supplementary Information for details)

The as-grown samples exhibit a low temperature ($T$< 12 K) upturn that is attributed to the logarithmic quantum interference corrections to the Drude conductivity $\Delta\sigma \approx \ln T$ that are characteristic in 2D systems (e.g. WL, WAL and electron-electron interaction (EEI)).[40, 41, 43-48] The resistivity is expected to decrease in the presence of WAL; therefore the observed increase in the resistivity with decreasing temperature in Figure 4a can only be associated with WL or EEI. Furthermore, in the case of TIs in the absence of backscattering, quantum interference effects lead only to WAL and EEI in systems with strong spin-orbit interactions.[16, 20] Therefore, this upturn in resistivity at reduced temperature ($T < 12$ K) in Figure 4a is likely due to EEI effects. Figure 4b illustrates the logarithmic temperature-dependent conductivity $\sigma(T)$ curves for all four samples. All curves are normalised by their maximum conductivities (e.g. the conductivity at 12 K). The linear slope of these curves are extracted and can be defined as $\kappa = (h/e^2)(\partial\sigma/\partial lnT)$ which is given in the legend of Figure 4b. A steeper slope indicates a larger $\kappa$ which means greater EEI.[39, 44, 49] The thicker samples (70 nm) show a higher $\kappa$ compared to the thinner samples (23 nm) which indicates a stronger EEI correction to the Drude conductivity at low temperature. A similar thickness dependence effect has been observed recently in polycrystalline $Bi_2Se_3$.[37]

**Discussion**

EEI and WAL are competing effects; at low temperatures, EEI will result in a decrease in conductivity whereas WAL produces a conductivity increase. So, it follows that the porous samples with an enhanced WAL exhibit a reduction in the EEI correction compared to the as-grown samples. The EEI reduction here ($\kappa$ reduction in figure 4b) is comparable to an experimental study where the EEI is reduced after fabricating nanostructured $Bi_2Te_3$ thin films.[50]

The most surprising result of this study is the fact that the phase coherence length doubled in the porous samples. Porosity increases the surface to bulk ratio, meaning the surface states contribute a larger fraction to the overall magnetotransport. The fact that the longitudinal resistivity decreases (compared to the as-grown) and is metallic in the porous samples indicates that the transport is primarily surface-like. Given the linear dispersion of Dirac states on the surface, the Fermi velocity ($v_F$) is expected to be large in comparison to the bulk since it is proportional to the slope of energy (E) versus **k** in momentum space.[51] It is important to note that the phase coherence length is proportional to $v_F$, that is $L_\varphi = \sqrt{D\tau_\varphi}$ and $D = v_F^2 \tau/d$, where $\tau_\varphi$ is the electron dephasing time, D is the electron diffusion constant, $\tau$ is the electron elastic mean free time and d is the system dimension.[52] Therefore, the large $v_F$ of the surface states could result in the enhanced $L_\varphi$ observed in the porous samples.

Our results indicate that the introduction of porosity produces disorder but does not destroy the topological surface states in the material. On the contrary, the phase coherence length nearly doubled. Such a result is promising for practical integration of topological insulators into future low energy electronic and spintronic devices, where disorder can be very difficult to eliminate entirely. Moreover, work has predicted that the porosity in a topological insulator can produce a large thermoelectric figure of merit due to the high contribution of the metallic surface states and suppressed phonon thermal conductivity.[27] Our work has experimentally shown that it is indeed possible to introduce porosity while still maintaining the metallic surface states, pointing to the promise of porous TIs in thermoelectric devices.

**Conclusion**

Temperature dependent resistivity and magnetotransport measurements were conducted on as-grow and porous $Bi_2Te_3$ thin films. Ion irradiation was used to introduce nanoscale porosity in the material, which increased the surface to volume ratio. The temperature dependence of the longitudinal resistivity indicates the porous films exhibit metallic-like behaviour owing to the increased surface state contribution. The EEI correction to the conductivity is present in all samples but stronger in the as-grown films. By contrast, the porous TIs exhibit stronger WAL and surprisingly a longer phase coherence length compared to the as-grown film. The larger phase coherence length is attributed to the large $v_F$ of the Dirac surface states. This work suggests porosity is an effective means to enhance the topological surface states' contribution to the transport, making these materials promising for low energy electronics, spintronics and thermoelectrics.

**Experimental Procedure**

Nanocrystalline $Bi_2Te_3$ thin films were grown at room temperature on amorphous $SiN_x$ on Si substrates using molecular beam epitaxy in an ultra-high vacuum (UHV) environment with a base pressure as low as $10^{-10}$ mbar. For $Bi_2Te_3$ film growth, effusion cells were used to evaporate elemental Bi (99.999%) and Te (99.95%) while the sample's temperature was kept at room temperature. Rates were calibrated with a quartz crystal microbalance before the growth. The $Bi_2Te_3$ thin films were grown using a 2:3 ratio of flux rates of Bi and Te. The films studied in this work were 23 nm and 70 nm thick.

To enable transport measurements, the films were grown through a shadow mask in the shape of a Hall bar with a channel length and width of 1 mm and 0.2 mm, respectively (Figure 1a). Ion irradiation was carried out using a 40 keV beam of Ne ions. The flux of ions irradiated on the sample was set to $1\times10^{16}$ ions per $cm^2$. Films were characterized before and after irradiation using JEOL JEM 2010 High Resolution Transmission Electron Microscopy (HRTEM) and a JEM-F200 Multi-purpose Electron Microscope. Raman and Energy-dispersive X-ray spectroscopy (EDS) was also conducted to examine the structure and composition of the cross-section. Temperature (2-300K) and magnetic field-dependent (±2T) longitudinal resistivity and temperature-dependent Hall effect measurements were performed in a Quantum Design Physical Properties Measurement System. Both AC and DC measurements were performed in constant current mode with an applied current of 350-400 nA and 3-5 µA, respectively. The AC measurements used a lock-in technique at 17 Hz. The same results were obtained from AC and DC measurements, as verified by measurements conducted on the same sample using the two techniques.

**Supporting Information**

Technical information and additional figures on HRTEM, EDX, Raman, EEI, Hall Effect, charge puddling, magnetoresistance and fitting the magnetoconductivity data to HLN formula.

**Corresponding Authors**
*gol.akhgar@monash.edu
*julie.karel@monash.edu

**ACKNOWLEDGMENT**

A.N., G.A., J.K., M.T.E., D.L.C., D.C., A.R.H and M.S.F., acknowledge the funding support from the Australian Research Council Centre for Excellence Future Low Energy Electronics Technologies (CE170100039). J. K., acknowledges the support from the Australian Research Council Discovery Projects (DP200102477 and DP220103783). This work was performed in part at the Melbourne Centre for Nanofabrication (MCN) in the Victorian Node of the Australian National Fabrication Facility (ANFF). The ion irradiation of this research was undertaken in Australian Neutron Science and Technology Organisation (ANSTO). We would like to acknowledge Prof. Steven Prawer and use of his laser cutter facility at the University of Melbourne in making the shadow masks.

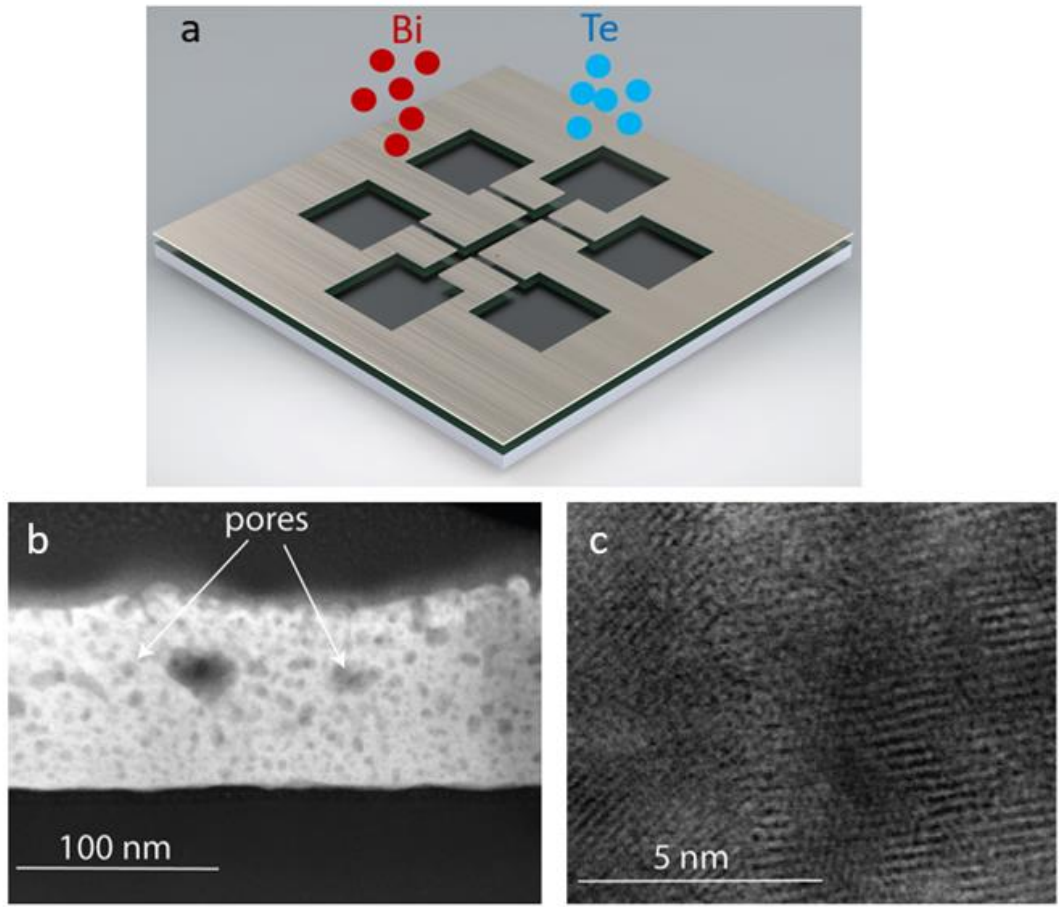

Figure 1. (a) Bi$_2$Te$_3$ growth through a shadow mask with Hall bar geometry onto a SiN$_x$ substrate with a channel length and width of 1 mm and 0.2 mm, respectively (b) Low magnification high angle annular dark field (HAADF) cross-sectional TEM image of the 70 nm thick irradiated Bi$_2$Te$_3$ Hall bar. The cross-section was prepared after transport measurements were conducted. The irradiated Ne ions produced pores (dark) dispersed throughout the Bi$_2$Te$_3$ thin film (light). (c) High magnification bright field image of the same sample depicting the nanocrystalline structure.

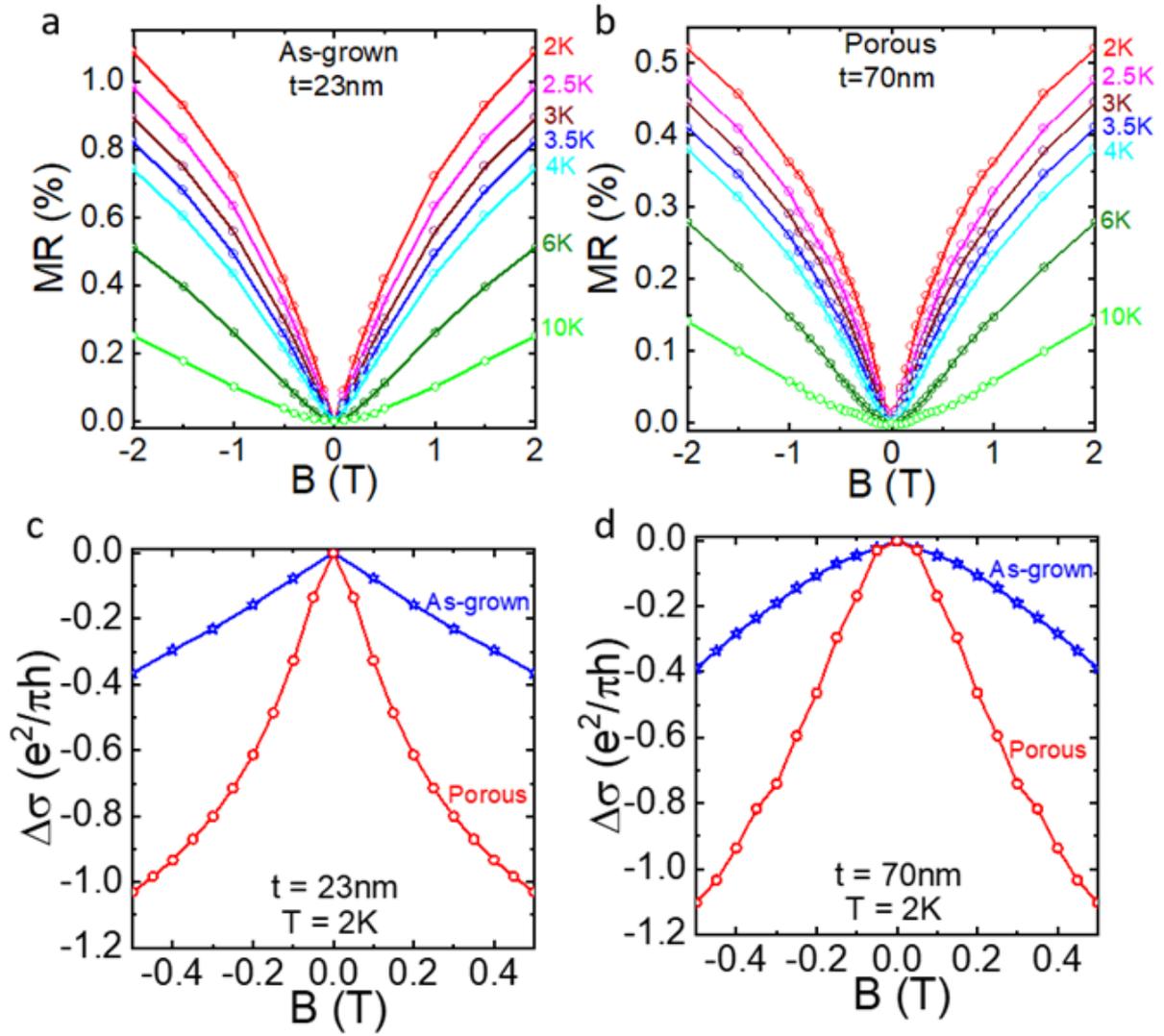

Figure 2: (a-b) Comparison of 2D magnetoconductivity of as-grown and porous samples with thickness of 23nm and 70nm respectively at 2K. (c) and (d) Normalised magnetoresistance % ($MR = (\rho_{xx}(B) - \rho_{xx}(0)) \times 100/\rho_{xx}(B = 0)$) as a function of magnetic field for the 23 nm as grown and porous $Bi_2Te_3$ respectively at different temperatures.

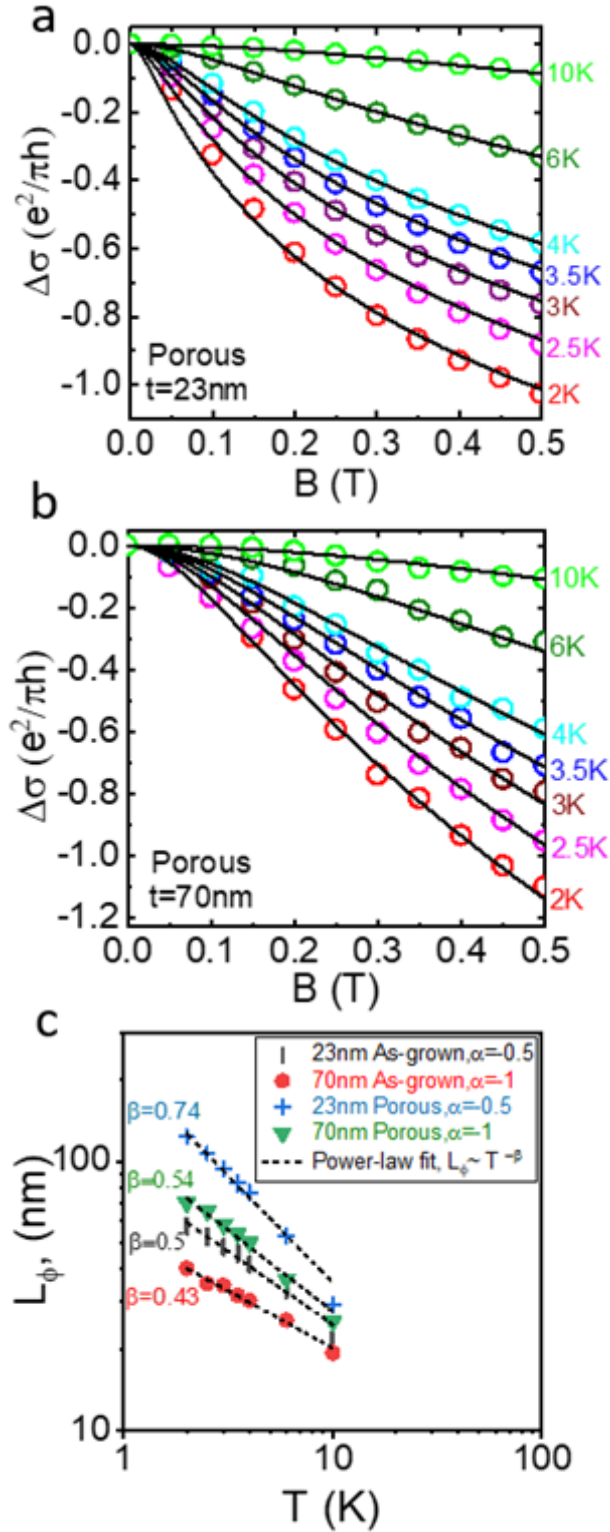

Figure 3: (a) 2D magnetoconductivity of porous $Bi_2Te_3$ with thickness of 23 nm and (b) 70 nm at various temperatures. The solid black line is the fit to the reduced HLN formula. (c) The temperature dependence of the phase coherence length of 23 nm as-grown (black vertical lines), 70 nm as-grown (red circles), 23 nm porous (blue crosses) and 70 nm porous (green triangles) $Bi_2Te_3$ samples in log-log scale. The dashed line is the fit to the power law, $L_\phi = T^{-\beta}$.

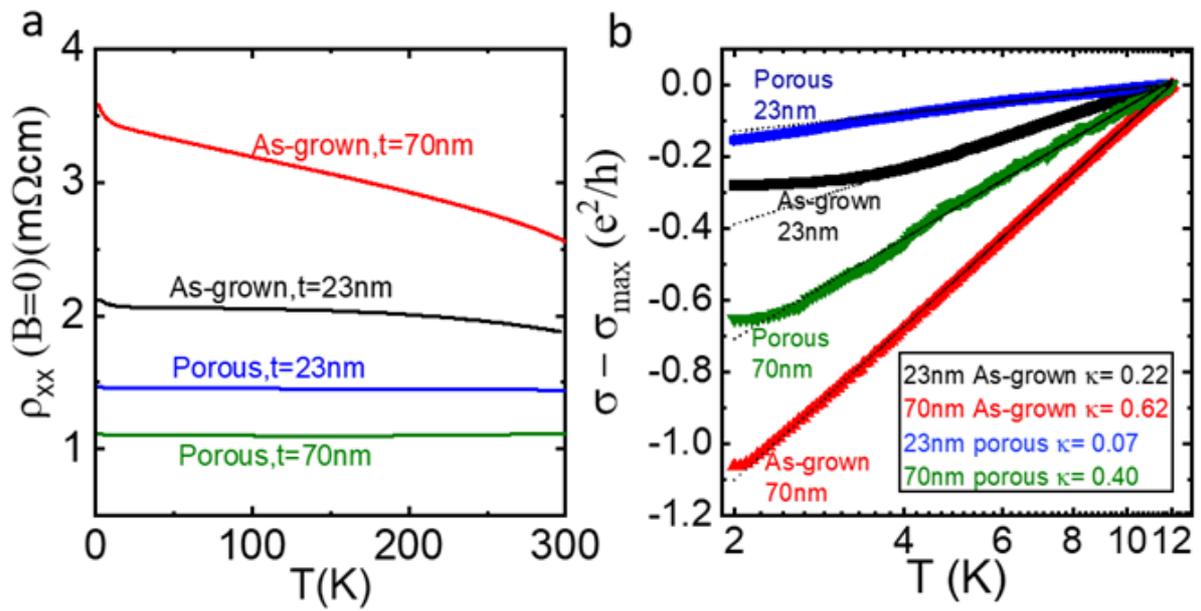

Figure 4: (a) Temperature dependence of resistivity and (b) logarithmic temperature-dependent conductivity at zero magnetic field for as-grown 23nm (black line), as-grown 70nm (red line), porous 23 nm (blue line) and porous 70 nm (green line) samples. The black dashed lines in (b) are the linear fit to the data.

# Supplementary Information

**Increased Phase Coherence Length in a Porous Topological Insulator**

Alex Nguyen[1,2]†, Golrokh Akhgar[1,2]*†, David L. Cortie[2,3,4], Abdulhakim Bake[3], Zeljko Pastuovic[4], Weiyao Zhao[1,2], Chang Liu[1,2], Yi-Hsun Chen[2,5], Kiyonori Suzuki[1], Michael S. Fuhrer[2,5], Dimitrie Culcer[2,6], Alexander R. Hamilton[2,6], Mark T. Edmonds[2,5] Julie Karel[1,2]*

[1]Department of Materials Science and Engineering, Monash University, Clayton, 3800, Victoria, Australia.
[2]ARC Centre of Excellence in Future Low-Energy Electronics Technology, Monash University, Clayton, 3800, Victoria, Australia.
[3]Institute for Superconducting and Electronic Materials, University of Wollongong, North Wollongong, 2522, New South Wales, Australia.
[4]Australian Neutron Science and Technology Organisation, New Illawarra Road, Lucas Heights NSW 2234, New South Wales, Australia.
[5]School of Physics and Astronomy, Monash University, Clayton, 3800, Victoria, Australia.
[6]School of Physics, The University of New South Wales, Sydney, 2052, New south Wales, Australia.

†These authors contributed equally to this work.
*Corresponding authors: gol.akhgar@monash.edu, julie.karel@monash.edu

## Size of Nanocrystals in Porous samples

High-resolution transmission electron microscopy (HRTEM) image from the as-grown films indicated a nanocrystalline structure, with an average crystal size of ~20nm. HRTEM was also conducted on porous samples. Figure S1a depicts nanocrystals and regions of disorder in the porous thin films. The nanocrystals in the porous sample were irregularly shaped, a sharp contrast with the nanocrystals in the as-grown films which had round edges and were generally oval in shape. Ion irradiation deformed the ovular shapes and created regions of disorder. Figure S1b illustrates a histogram of the longest dimension of the nanocrystals in the TEM images. None of the observed nanocrystals in the areas imaged had a dimension above 10 nm which suggests that introducing porosity may have reduced the size of the nanocrystals.

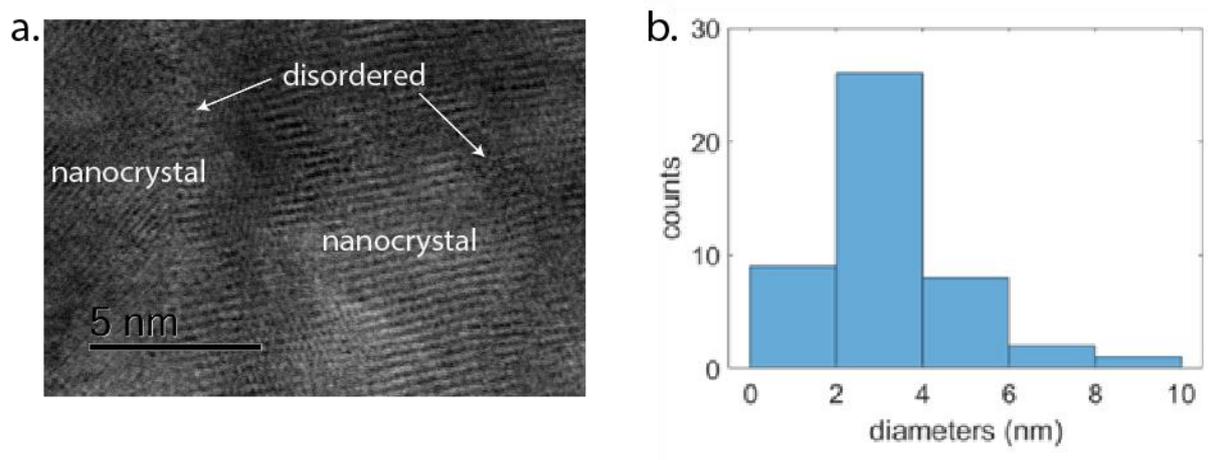

Figure S1: (a) A TEM image of a nanocrystals and disordered regions in the 70 nm porous $Bi_2Te_3$ thin films. (b) A histogram of the longest dimension of various nanocrystals in the porous thin films.

## Pore Characterisation

Figure S2a is a cross-sectional HRTEM image of the Hall bar. This image was used as a sampling area to measure the size of the cavities in the thin films. A majority of the pores (99%) were between 1 and 10 nm in size whereas the remaining voids were outliers with diameters over 30 nm large. The pores were assumed to be spherical in order to determine the total surface area. It was concluded that about 22% of the surface area was composed of pores.

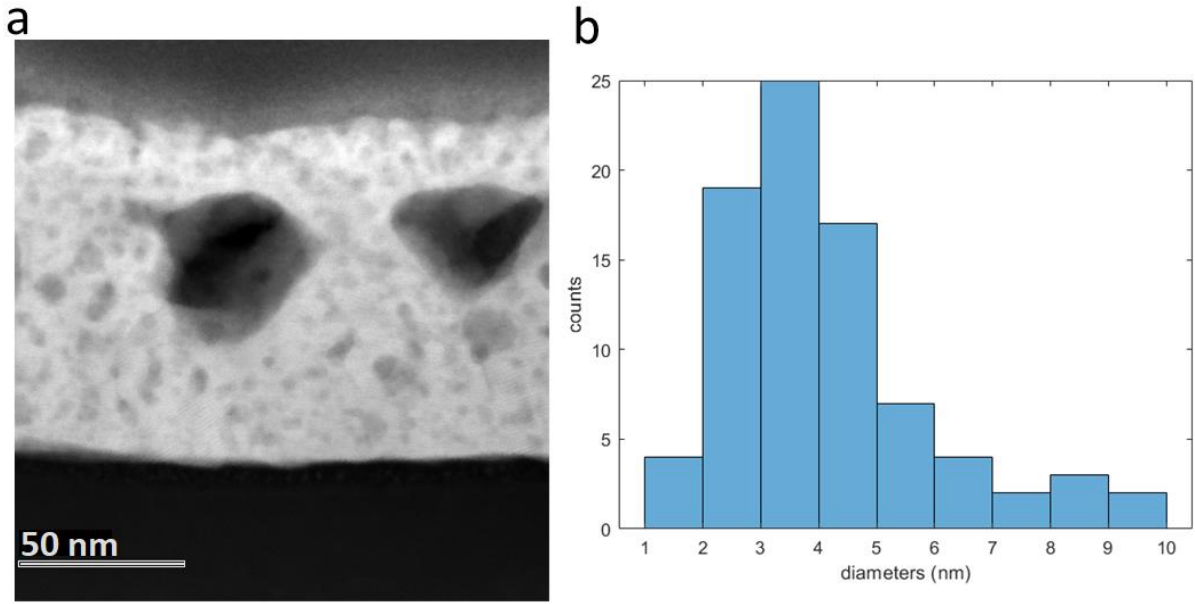

Figure S2: (a) A HRTEM image of a cross sectional area of the Hall bar. The size of the cavities in this region was measured. (b) A histogram illustrating the size distribution of the small holes. The size of the two largest holes are omitted but they were both over 30 nm large.

## Raman Spectroscopy

A home-made micro-Raman spectrometer (Acton SpectraPro SP-2750, Princeton Instruments) equipped with optical access was used to measure Raman spectra on epitaxially grown $Bi_2Te_3$ and as-grown $Bi_2Te_3$ samples. The laser beam (Sapphire SF at 532 nm) is focused on the sample by a 100× objective lens (NA: 0.8) and reflected by a nonpolarizing cube beam splitter into the spectrometer. The laser power is fixed at 0.35 mW during the entire measurement. Figure S3 depicts the Raman spectra of both samples in the region of 50-200 cm$^{-1}$. The two main peaks were observed and labeled as $E_g^2$ and $A_{1g}^2$. $E_g^2$ and $A_{1g}^2$ peaks correspond to lattice vibrational modes and are very close to the previously measured Raman peaks of the $Bi_2Te_3$.[1-3] $E_g^2$ and $A_{1g}^2$ modes are at 101 cm$^{-1}$ and 132 cm$^{-1}$ respectively for epitaxial grown $Bi_2Te_3$ sample (black line) and at 101 cm$^{-1}$ and 128 cm$^{-1}$ respectively for the as-grown $Bi_2Te_3$ sample (red line). The blue dash line is the full width half maximum (FWHM) fit to the data, and it shows the Raman peaks to be broadened in the as grown $Bi_2Te_3$ sample compared to epitaxial grown sample. The FWHM of the $E_g^2$ peak is broadened from ~5.1 cm$^{-1}$ to ~9.9 cm$^{-1}$. This confirms that the spatial vibration is reduced by introducing disorder in the as-grown $Bi_2Te_3$. The local bonding is still maintained. The broadening effect in the as-grown sample could also explain the slight shift of the pronounced $A_{1g}^2$ mode by approximately 4 cm$^{-1}$ in wavenumber as Van de Waals (VdW) layers are decreased.

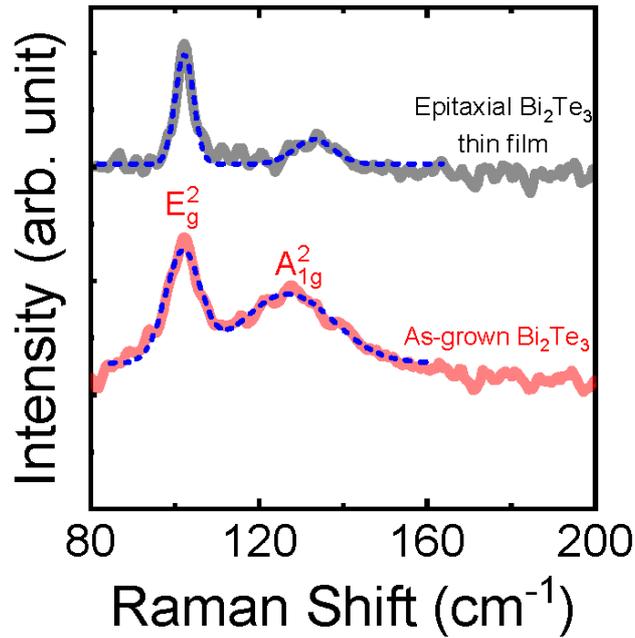

Figure S3: Raman spectra of epitaxial grown $Bi_2Te_3$ thin film (black line) and as-grown $Bi_2Te_3$ (red line). Blue dashed line is the full width half maximum (FWHM) fit to the data.

**Energy-dispersive X-ray spectroscopy (EDS)**

Figure S4 shows the EDS results on the 70 nm porous $Bi_2Te_3$ sample. No Ne ions were detected after ion irradiation. Furthermore, the Bi/Te ratio in the table below shows that the Bi/Te composition deviates from the nominal composition due to differential surface tension effects and constitutional supercooling effects in $Bi_2Te_3$.[1] Therefore, Te has mostly coalesced at the $Bi_2Te_3$/substrate interface which has led to measurements of Te poor $Bi_2Te_3$ away from the interface.

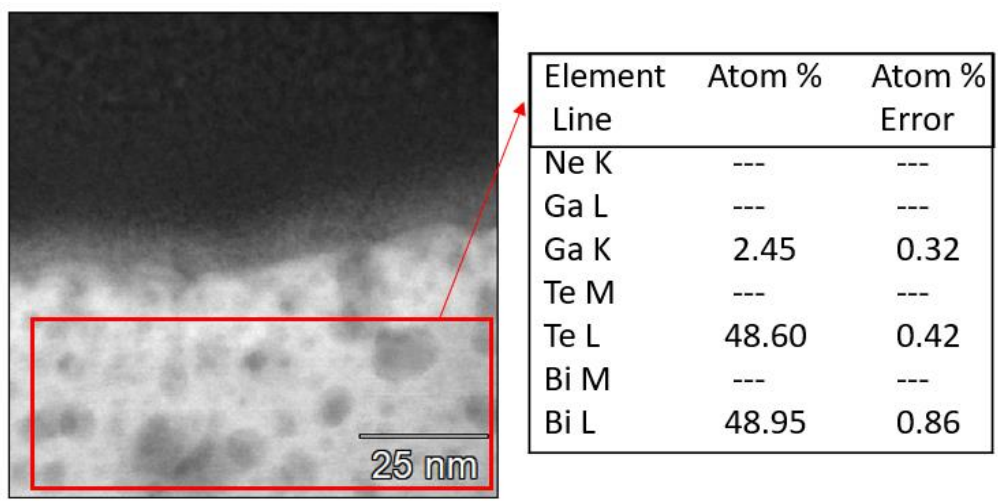

Figure S4: EDX measurement on the 70 nm Porous $Bi_2Te_3$ sample.

# Hall Effect Measurements and Charge Puddling

The Hall effect data can be used to extrapolate the sign of the charge carriers and the number density of charge carriers.[4] Figure S5 shows the transverse magnetoresistivity of (a) 23 nm as-grown, (b) 70 nm As-grown, (c) 23 nm porous and (d) 70nm porous thin films. The 70 nm porous sample appears to have holes as charge carriers whereas other samples appear to have electrons as charge carriers. This apparent change in charge carrier type is regularly observed in crystalline TIs and attributed to charge puddling in the material.[5]

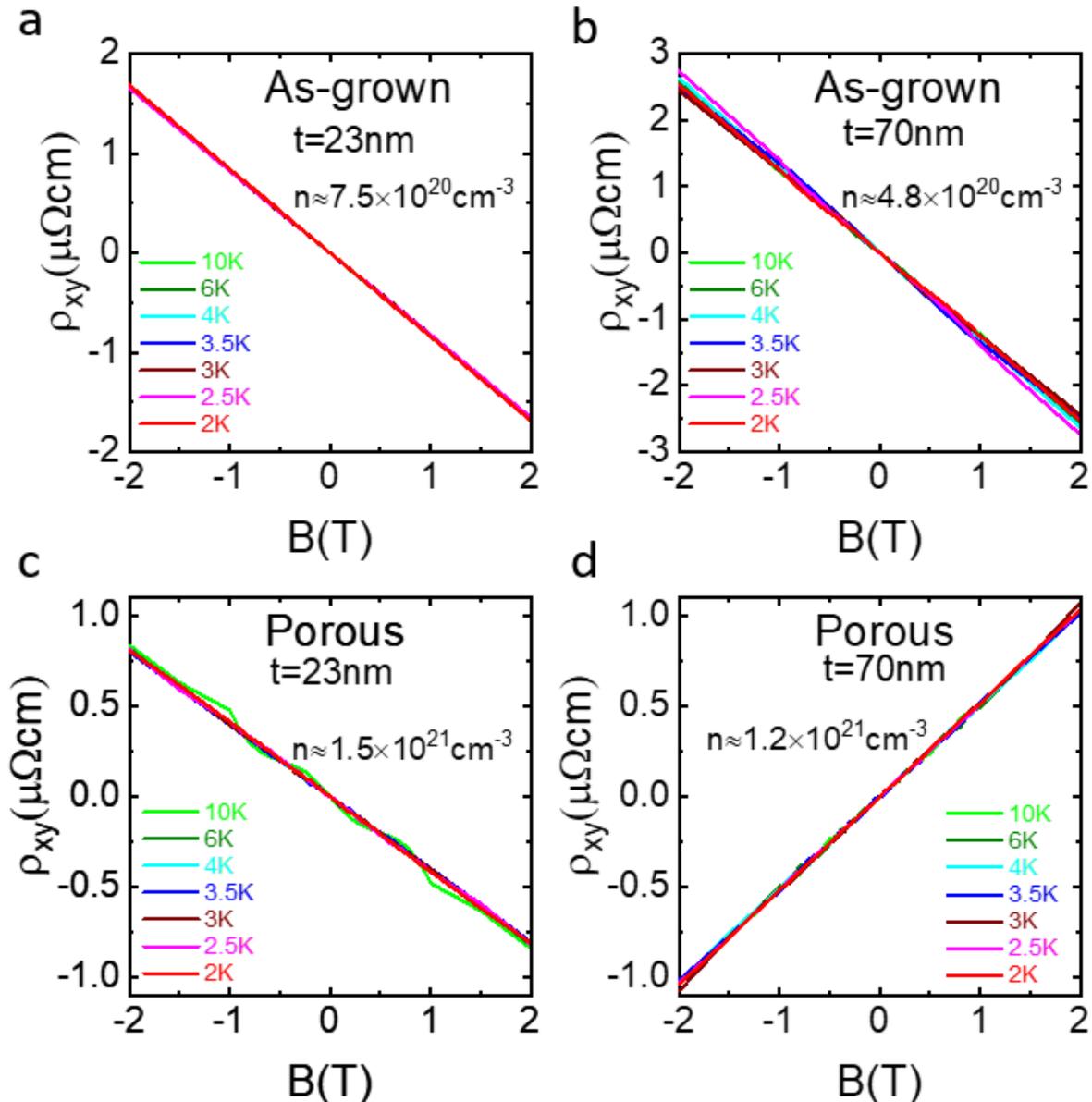

Figure S5: A) Hall resistivity as a function of magnetic field of (a) 23 nm As-grown, (b) 70nm As-grown, (c) 23nm Porous and (d) 70nm Porous Hall bars. *n* refers to the number density of charge carriers.

In order to confirm charge puddling, we have performed an analysis of the magnetoresistivity curves.[6-8] At high temperature and large magnetic field, the magnetoresistivity curve should be parabolic i.e.:

$$\frac{\rho_{xx}(B) - \rho_{xx}(B=0)}{\rho_{xx}(B=0)} = A\mu^2 B^2 \tag{1}$$

where $\rho_{xx}$ is the longitudinal resistivity, B is the magnitude of the applied magnetic field, $A$ is a dimensionless parameter and $\mu$ is the mobility.[6-8] $A$ is theoretically predicted to be between 0 and 0.5. $A=0.5$ is associated with 2D transport.[6] However, $A$ has been experimentally observed between 0 and 1.[6,7] The magnetoresistivity of the 23 nm porous sample at 20 K is plotted in Figure S6. Specifically, the magnetoresistivity was plotted against $B^2$. The high field data was then fitted to a parabola (equation 1). The dashed line is the linear fit to the data. Setting $A$ in equation 1 to be 0.5 the mobility can be extrapolated to be about 140 cm$^2$V$^{-1}$s$^{-1}$. Additionally, a number density of charge carriers was extrapolated from the Hall effect data as illustrated in Figure S5c. The Drude model states that $\sigma = e\mu n$ where $\sigma$ is the 20 K conductivity.[4] Using the charge carrier number density extrapolated from the Hall effect, the mobility can be calculated as 3 cm$^2$V$^{-1}$s$^{-1}$; this mobility will be referred to as the Hall mobility. Inserting the Hall mobility in equation 1, $A$ can be determined as 1052, an unrealistic value. This unrealistic value for $A$ and the disagreement of the two mobilities indicates charge puddling is present.[6-8] Additionally, charge puddling causes the Hall mobility to underestimate the actual mobility. Therefore, the very low Hall mobility can also be understood as a result of charge puddling.

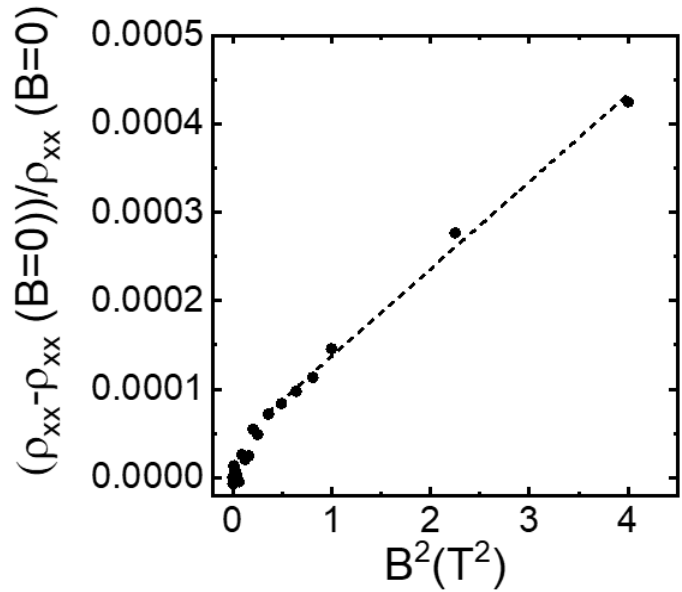

Figure S6: A plot of the normalized magnetoresistivity of the 23 nm porous sample at 20 K. The data points collected at high field were fitted to equation 1. The fit is illustrated by the dashed line. The x-axis depicts the applied magnetic field squared. This was done to simplify the quadratic fit into a linear fit.

## Two-dimensional resistivity

Figure S7 is showing the 2D sheet resistivity as a function of temperature. This figure shows that the 2D conduction has enhanced after ion irradiation in porous samples and as results the resistivity is reduced in both thicknesses.

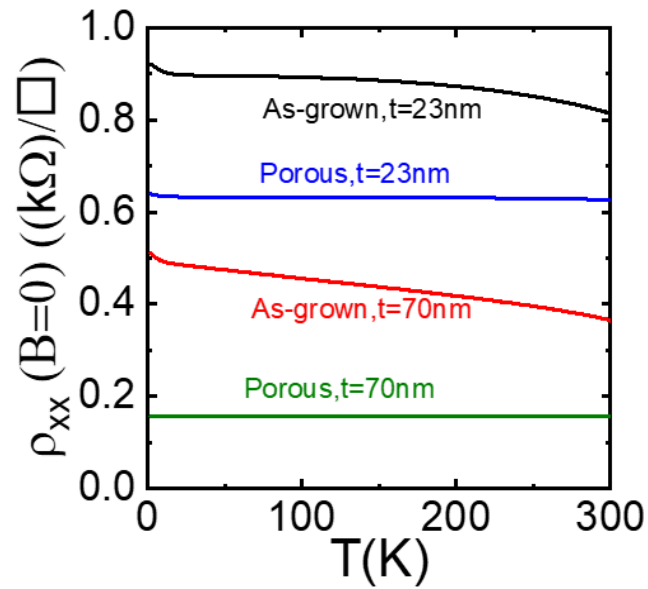

Figure S7: 2D sheet resistivity as a function of temperature for as-grown 23nm (black line), as-grown 70nm (red line), porous 23 nm (blue line) and porous 70 nm (green line) samples.

# Logarithmic Temperature-dependent Conductivity

Figure S8 a and b are semilogarithmic σ(T) plots at different applied magnetic fields for the 23 nm as-grown and 23 nm porous samples respectively. These figures show that by increasing the applied magnetic field the slope becomes steeper, therefore, κ increases. Figure S8c is showing the extracted κ for the 23 nm as-grown and porous samples as a function magnetic field. In both samples, the slope saturates above 0.5 T, indicating the localisation tendency is dominated by the applied magnetic field. The behaviour of κ and slope reduction in the porous sample compared to as-grown is comparable to an experimental study where the EEI is reduced after fabricating nanostructured in $Bi_2Te_3$ thin films.[9]

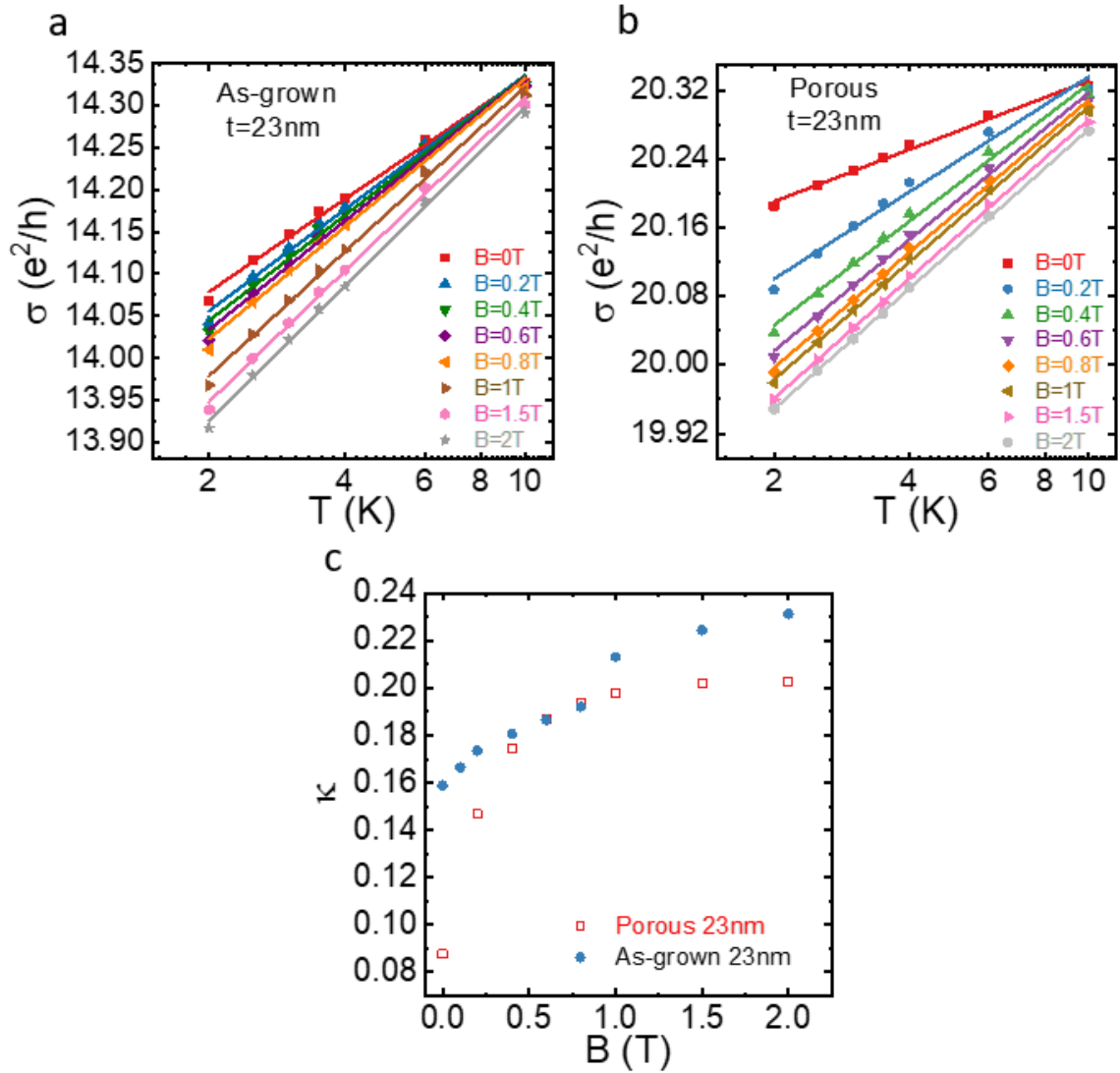

Figure S8: Logarithmic temperature-dependent conductivity of 23 nm (a) As-grown and (b) porous samples at different applied magnetic field. Solid lines are the fit to the data. (c) Slope of semilogarithmic σ(T) curve, κ as a function of magnetic field B for porous and as-grown samples.

## Magnetoresistance

Figure S9 a-d shows the magnetoresistance at various temperatures, as indicated, for all samples.

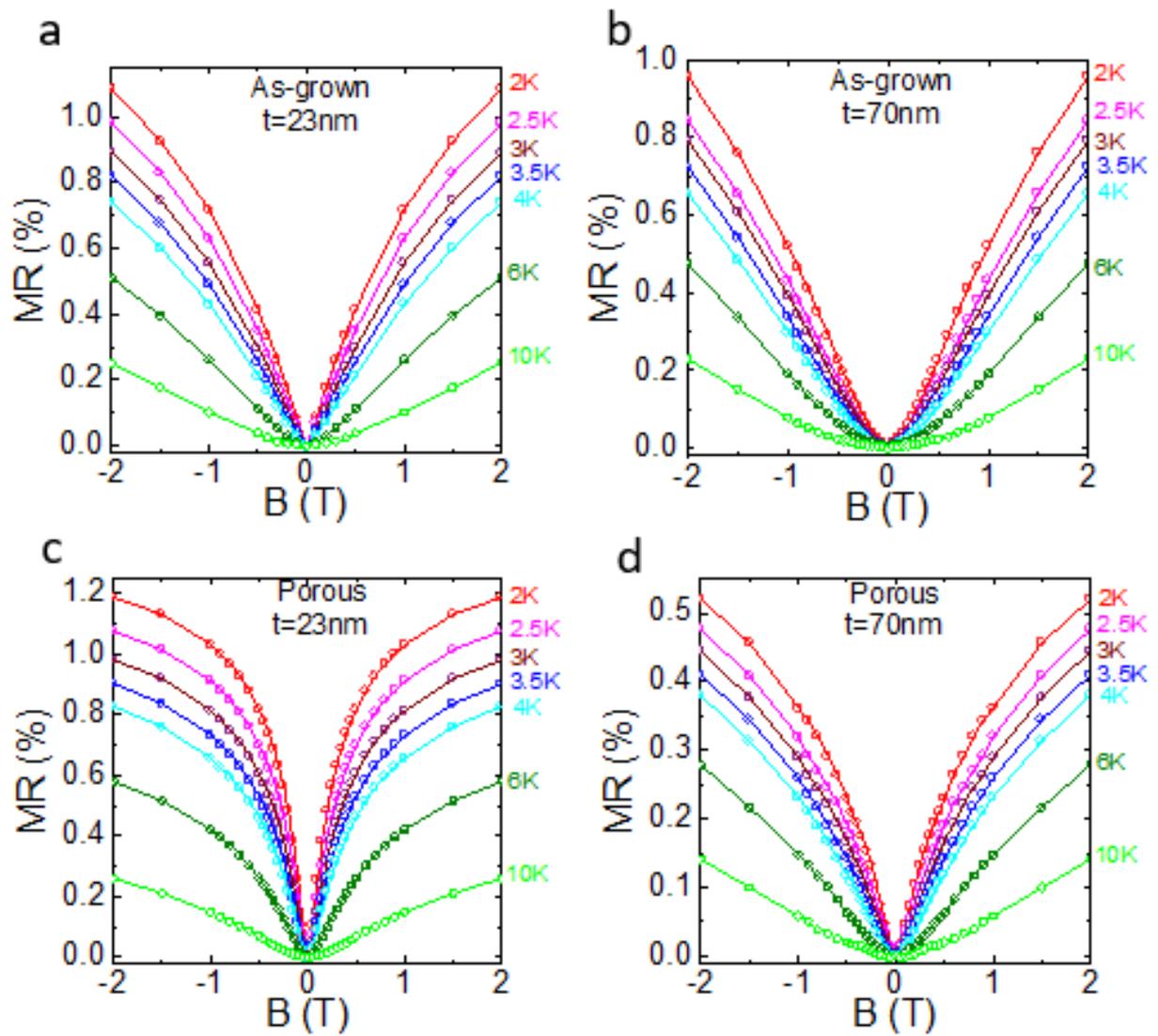

Figure S9: Magnetoresistance of (a) 23 nm As-grown, (b) 70 nm As-grown, (c) 23 nm Porous, and (d) 70 nm Porous samples.

## Different Thickness Magnetoconductivity Comparisons

Figure S10 shows a comparsion of the magnetoconductivity for (a) the as-grown and (b) the porous samples of different thicknesses, as indicated.

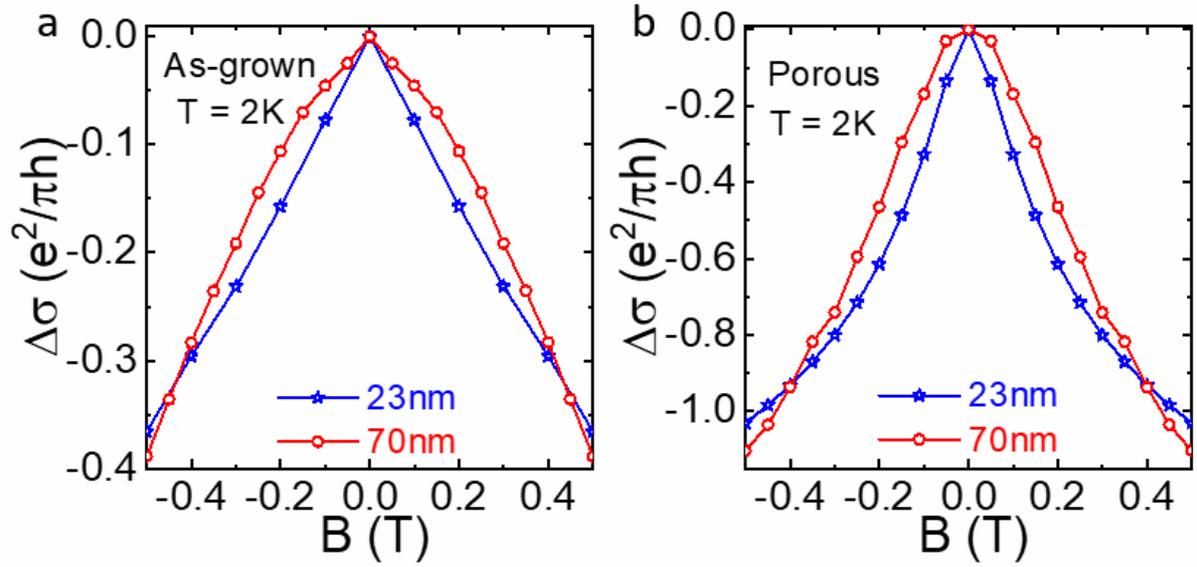

Figure S10: 2D magnetoconductivity $\Delta\sigma_{xx} = \sigma_{xx}(B) - \sigma_{xx}(B=0)$ in the unit of $e^2/\pi h$ of (a) as-grown (b) Porous $Bi_2Te_3$ samples where the blue line is the 23nm and red line is the 70nm sample at 2 K

## Magnetoconductivity Fits to the HLN Theory

Figure S11 a-d shows the fits to the magnetoconductivity using the reduced HLN formula for all samples.

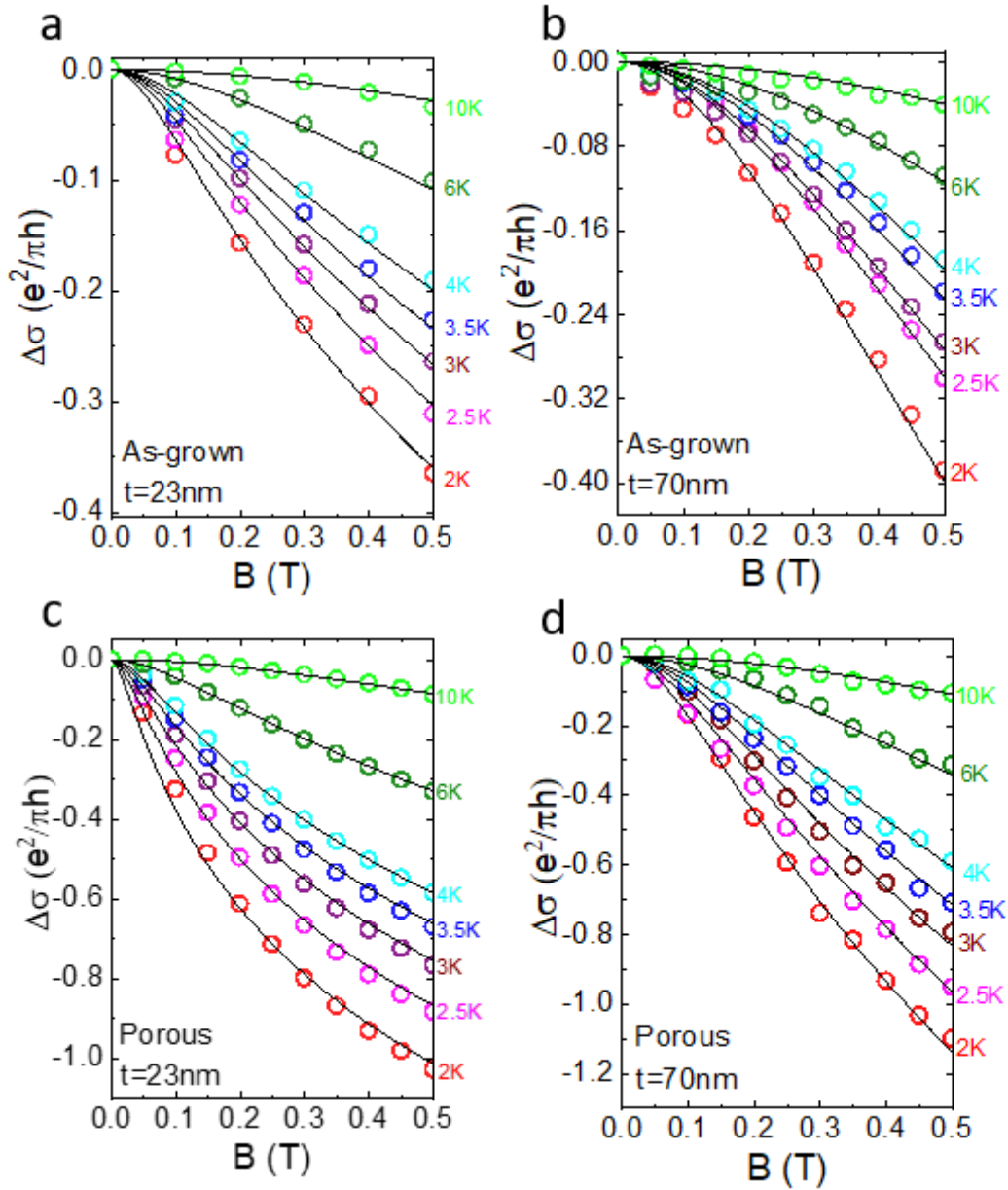

Figure S11: 2D magnetoconductivity of (a) 23nm As-grown, (b) 70nm As-grown, (c) 23nm Porous, and (d) 70nm Porous samples at various temperatures. The black solid line is the fit to the reduced HLN formula.